\documentclass[prl,preprintnumbers,twocolumn]{revtex4}

\usepackage{graphicx}
\usepackage{dcolumn}
\usepackage{bm}

\usepackage{amsmath}
\usepackage{amssymb}

\newcommand{\be}{\begin{equation}}
\newcommand{\ee}{\end{equation}}

\def\ph{\varphi }

\def\0{\over } \def\2{{1\over2}} \def\4{{1\over4}}
\def\5{\hat } \def\6{\partial }

\def\del{\partial }

\begin{document}

\preprint{T03/007}

\title{Renormalizability of $\Phi$-derivable approximations in scalar
$\varphi^4$ theory}

\author{Jean-Paul Blaizot}
 \email[]{blaizot@spht.saclay.cea.fr}
\author{Edmond Iancu}
\email[]{iancu@spht.saclay.cea.fr}
\author{Urko Reinosa}
\email[]{reinosa@spht.saclay.cea.fr}
   \affiliation{Service de Physique Th\'eorique, CEA/DSM/SPhT,
91191 Gif-sur-Yvette Cedex, France.}

\date{\today}

\begin{abstract}
We discuss the renormalizability of $\Phi$-derivable approximations
in  scalar $\varphi^4$ theory in
four dimensions. The formalism leads to self-consistent equations for the
  2-point and the 4-point functions which
  are    plagued by ultraviolet divergences. Through a detailed analysis of
  the one and two-loop self-energy skeletons,  we show that both 
equations can be
renormalized simultaneously  and  determine the corresponding
counterterms. These insure the elimination  of ultraviolet
divergences both at zero and finite
temperature.
\end{abstract}

\pacs{Valid PACS appear here}
\maketitle

\newcommand \beq{\begin{eqnarray}}
\newcommand \eeq{\end{eqnarray}}
\newcommand{\ba}{\begin{eqnarray}}
\newcommand{\ea}{\end{eqnarray}}

\input epsf


\def\square{\hbox{{$\sqcup$}\llap{$\sqcap$}}}
\def\grad{\nabla}
\def\del{\partial}


\def\frac#1#2{{#1 \over #2}}
\def\smallfrac#1#2{{\scriptstyle {#1 \over #2}}}
\def\half{\ifinner {\scriptstyle {1 \over 2}}
     \else {1 \over 2} \fi}


\def\bra#1{\langle#1\vert}
\def\ket#1{\vert#1\rangle}


\def\simge{\mathrel{%
     \rlap{\raise 0.511ex \hbox{$>$}}{\lower 0.511ex \hbox{$\sim$}}}}
\def\simle{\mathrel{
     \rlap{\raise 0.511ex \hbox{$<$}}{\lower 0.511ex \hbox{$\sim$}}}}


\def\buildchar#1#2#3{{\null\!
     \mathop#1\limits^{#2}_{#3}
     \!\null}}
\def\overcirc#1{\buildchar{#1}{\circ}{}}


\def\slashchar#1{\setbox0=\hbox{$#1$}
     \dimen0=\wd0
     \setbox1=\hbox{/} \dimen1=\wd1
     \ifdim\dimen0>\dimen1
        \rlap{\hbox to \dimen0{\hfil/\hfil}}
        #1
     \else
        \rlap{\hbox to \dimen1{\hfil$#1$\hfil}}
        /
     \fi}

Self-consistent, ``$\Phi$-derivable'', approximations were introduced
many years ago in the
context of the non-relativistic many body problem
\cite{LW,Baym}, and have been extended to field theory
\cite{Cornwall:vz,Baym:qb}.
   They have been found appropriate  to
treat systems for which the quasiparticle picture is a good starting
point and  have recently been applied in this
spirit   to calculate equilibrium thermodynamics of the quark-gluon plasma
\cite{Blaizot:2000fc}. They are also being used to study the dynamics
of quantum fields out of equilibrium
\cite{Aarts:2002dj}.

   The main
difficulty in implementing  such approximations in quantum
field theory is their renormalization: from the point of view of
perturbation theory, the  equations that one is led to
solve effectively resum   infinite sets of Feynman diagrams, and the
existence of a procedure for constructing the
counterterms needed to eliminate the corresponding divergences is not
obvious. This problem becomes particularly acute
at finite temperature: While, on general grounds, one expects
ultraviolet divergences to be  unaffected
by the temperature (see e.g. \cite{Collins:xc}), in
self-consistent approximations temperature dependent
divergences often do appear, thus casting doubts on the
renormalizability (see in particular
\cite{Braaten:2001vr}).

This issue has been addressed recently by van Hees and Knoll in a
series of papers
\cite{vanHees:2001ik,VanHees:2001pf}.
The strategy put forward in \cite{vanHees:2001ik} is based on an
expansion of the propagator around the vacuum
self-consistent solution, and  relies
on the real time formalism. The
elimination of the divergences proceeds through the BPHZ  subtraction
scheme. This  leads to  a
systematic and practical renormalization scheme where  temperature
dependent counterterms never appear. However the
dissymmetrical treatment of  the vacuum sector and the finite
temperature one is unsatisfactory: It hides the
fact that the rearrangement  of  divergences which appears to be
necessary at finite temperature is
also needed in most renormalization schemes already at zero temperature.
And it does not bring out the specific relation between the bare
and the renormalized parameters
that emerges in $\Phi$-derivable approximations. 
This makes it difficult, e.g., to compute the $\beta$-function,
or resolve the apparent discrepancy between
the results of Refs. \cite{Braaten:2001vr} and \cite{vanHees:2001ik}. 

We have therefore reconsidered the problem from a more general
perspective.
Our derivations  use the imaginary time formalism, making the
connection with conventional equilibrium field
theory transparent, and allowing for a simultaneous treatment of the
vacuum sector and the finite temperature one:
once renormalization is done properly at zero temperature, the
extension to finite temperature is straightforward.

The central quantity in $\Phi$-derivable approximations is $\Phi[D]$,
the sum of the
2-particle-irreducible ``skeleton'' diagrams, a  functional of the
full propagator $D$, which enters
the expression of the thermodynamical potential. From $\Phi[D]$ we
may calculate the 2-point function (the self-energy)
by functional differentiation:
\beq\label{PhiPi}
\delta \Phi[D]/\delta D\,=\, \frac{1}{2}\,\Pi\,.
\eeq
This relation, together with
Dyson's equation ($D_0$ denotes the bare propagator):
\beq\label{Dyson}
D^{-1}=D^{-1}_0+\Pi[D],
\eeq
defines the
physical propagator and self-energy in a self-consistent way. We
shall refer to Eq.~(\ref{Dyson}), with $\Pi$ given by
(\ref{PhiPi}), as the ``gap equation''.  A further differentiation of
$\Phi$ with respect to $D$
yields the 2-particle irreducible kernel
\beq
\Lambda(K,P)=2\frac{\delta \Pi(K)}{\delta D(P)}=\Lambda(P,K)
\eeq
of
a Bethe-Salpeter (BS) equation
\begin{equation}\label{BS1}
\Gamma(K,P)=\Lambda(K,P)-\frac{1}{2}\int_Q \Gamma(K,Q)D^2(Q)\Lambda(Q,P)
\end{equation}
that allows the calculation of the 4-point function $\Gamma(K,P)$ with a degree
of accuracy comparable with
that used in the determination of the propagator. $\Phi$-derivable
approximations
are  obtained by selecting a class of skeletons in
$\Phi[D]$ and calculating $\Pi$ and $\Gamma$ from the equations above.
As we shall see, the renormalizability of such approximations relies
on the possibility to simultaneously
renormalize
$\Pi$ and
$\Gamma$. In particular, the BS equation is needed to determine
coupling constant counterterms which
eliminate some divergences of the self-energy.

\begin{figure}
\includegraphics[width=5cm]{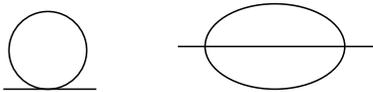}
\caption{The one-loop and two-loop skeleton diagrams contributing to
the self-energy. These will be referred to as the
``tadpole'' and ``sunset'' diagrams, respectively.}
\end{figure}

We consider in this paper a massive scalar field theory with a
$\varphi^4$ interaction:
\beq
\mathcal{L}=\frac{1}{2}\left(\partial\ph_0\right)^2-\frac{1}{2}m_0^2\ph_0^2-
\frac{1}{4!}g_0^2\ph_0^4,
\eeq
and include in  $\Phi$ only the 2-loop and 3-loop  skeletons (the
corresponding self-energy diagrams are
displayed in Fig.~1). This allows us to introduce the generic
difficulties, deferring the
systematic discussion of the general case to a forthcoming publication
\cite{BIU03}.
In four dimensions, usual power counting indicates that only the 2-point
     and the 4-point functions are divergent. The divergent parts can
be absorbed in local countertems
corresponding to a redefinition of the parameters of the  lagrangian.
We assume standard relations \cite{Collins:xc} between the
renormalized and bare parameters: $\ph_0=\sqrt{Z}\ph$,
$Zm_0^2=m^2+\delta m^2$, $Z^2 g_0^2=g^2+\delta g^2$, and $\delta
Z=Z-1$.

The  gap equation corresponding to the one-loop skeleton (the
``tadpole'') reads:
\beq\label{Pitad}
\Pi = \frac{g^2}{2}\int_P D(P) +\delta m^2
\eeq
where $D^{-1}(P)=P^2+m^2+\Pi $. The notation $\int_P$ stands for an
Euclidean integral over the 4-momentum $P$. At
finite temperature, it should be understood as an integral over the
3-momentum together with a sum over
Matsubara frequencies. The approximation corresponding to
Eq.~(\ref{Pitad}) is a simple
self-consistent mean field approximation that has been treated many
times before (see for instance
\cite{Gross:jv,Dolan:qd,Baym:qb,Drummond:1997cw}). We  present it
here in a way which will prepare for the
more complicated two-loop example that we shall discuss next.

The self-energy
$\Pi$ is here a constant, and a single mass counterterm $\delta m^2$
is in principle sufficient to eliminate the
ultraviolet divergence.  Calculating the integral in
Eq.~(\ref{Pitad}) in dimensional regularization we get:
\beq\label{int1}
\mu^{2\epsilon}\!\!\int_P D(P)\!\!=\!-\frac{1}{16\pi^2}
(m^2+\Pi)\!\left\{ \frac{1}{\epsilon}
-\ln\frac{m^2+\Pi}{\bar\mu^2}+ 1\right\}
\eeq
where $\bar\mu^2\equiv 4\pi {\rm e}^{-\gamma_E}\mu^2$. At this point,
one could be tempted to absorb the whole
divergence in
$\delta m^2$, i.e., set:
\beq\label{dmnaive}
\delta m^2\,=\,\frac{g^2}{32\pi^2}
\,(m^2+\Pi)\, \frac{1}{\epsilon}\,.
\eeq
But this is not a good strategy. If, for instance, the calculation
is done at finite
temperature,
$\Pi$ depends on the temperature, and so does the  counterterm
(\ref{dmnaive}), which we want to avoid.

In fact, when analyzing the gap equation  Eq.~(\ref{Pitad}) in terms
of perturbation theory,  on finds that
its solution effectively resums an infinite set of Feynman diagrams,
some of which contribute to the
renormalization of the coupling constant. This is best seen by
imagining solving this equation  by iteration, a
   procedure which also defines an explicit construction of the
counterterms. To do so, we set
$D(P)=D_0(P)=(P^2+m^2)^{-1}$ in the r.h.s. of Eq.~(\ref{Pitad}); one
then  obtains a first approximation to $\Pi$ on
the l.h.s, which can  then be used in the r.h.s., and so on. At each
iteration, $\delta m^2$ can be adjusted to absorb
the
\textit{overall} divergence. But is is easy to see that, starting at
the second iteration, a \textit{subdivergence}
appears  corresponding to a coupling constant renormalization that
needs to be subtracted before adjusting $\delta
m^2$ (An illustration of the phenomenon in the less trivial example
of the sunset diagram is given in Fig.~2 below).
New such subdivergences  appear in each iteration, and to take them
into account,  a term of the form
$(\delta g^2/2)\int_P D(P)$ should be added in the r.h.s. of
Eq.~(\ref{Pitad}). Equivalently  $g^2$ should be replaced
by
$g_0^2=g^2+\delta g^2$ in Eq.~(\ref{Pitad}). As we shall see,
$\delta g^2$ is precisely the counterterm that is
needed to make  finite the BS equation, to which we now turn.

With, here, $\Lambda=g_0^2=g^2+\delta g^2$, the BS equation reads:
\beq\label{BStad}
\Gamma=g_0^2-\frac{g_0^2}{2}\,\Gamma\int_P
D^2(P),
\eeq
where $\Gamma$ is the renormalized 4-point function, and $\delta g^2$
is chosen so as to absorb the divergence of the
integral. Note that this divergence  does not depend on the mass (nor
therefore on $\Pi$), and
for the purpose of determining $\delta g^2$  we could as well use an
auxiliary 4-point function
$\Gamma_0$ solution of Eq.~(\ref{BStad}) with  $D$  replaced by
$D_0$. Clearly, $\Gamma_0$ differs from $\Gamma$ by a
finite quantity only.

We now return to the gap equation, Eq.~(\ref{Pitad}) with $g^2$
replaced by $g_0^2$, itself determined in terms of
$\Gamma_0$ by the BS equation,  and show that  its solution, $\Pi$,
can be made finite with a counterterm $\delta m^2$
independent of $\Pi$. To this aim, we write
$D=D_0+\delta D$, where
\beq
\delta D(P)=D_0(P)[-\Pi]D_0(P)+D_r(P),
\eeq
and $D_r(P)$ starts at order $\Pi^2$, so that the integral
$\int_P D_r(P)$ is finite. Then, we set
\beq
\tilde \Pi_2=\frac{g_0^2}{2}\!\int_P D_0(P)+\delta m^2\qquad \tilde
\Pi_0=\frac{g_0^2}{2}\!\int_P \delta D(P),\,\,
\eeq
where only $\tilde \Pi_0$ depends on $\Pi$, and when $\Pi$ is
solution of the gap equation, $\Pi=\tilde \Pi_0+\tilde
\Pi_2$. Next, one uses the BS
equation to eliminate $g_0^2$ in the defining equation for $\tilde\Pi_0$:
\beq\label{tildePi0tad}
\tilde \Pi_0=\frac{\Gamma_0}{2}\int_P \delta
D(P)+\frac{\Gamma_0}{2}\tilde\Pi_0\int_Q D_0^2(Q).
\eeq
At this point, we have achieved our goal: while both integrals in
Eq.~(\ref{tildePi0tad}) are divergent, it is easily
verified that no divergence involves $\Pi$  when $\Pi$ is solution of
the gap equation (so that we can replace
$\tilde\Pi_0$ by $\Pi-\tilde\Pi_2$ in the r.h.s. of
Eq.~(\ref{tildePi0tad})). The final gap equation reads:
\beq
\Pi=\frac{\Gamma}{2}\int_P
D_r(P)+\tilde\Pi_2\left(1-\frac{\Gamma}{2}\int_Q D_0^2(Q)\right).
\eeq
The divergence in the last term involves only $D_0$ and  can
be absorbed in $\delta m^2$.  For instance,
in dimensional regularization with minimal subtraction $\delta
m^2={g_0^2m^2}/({32\pi^2\epsilon})$. The factor
multiplying
$\tilde\Pi_2$ is
$\Gamma_0/g_0^2$ (from the BS equation), so that the resulting
expression is indeed finite (but depends on the
scheme).

At finite temperature, the one-loop integral in Eq.~(\ref{Pitad}) can
be split into a vacuum integral and a
3-dimensional
   integral involving a statistical factor and giving the following new
contribution to $\Pi$:
\beq\label{tildePi1}
\tilde\Pi_1=\frac{g_0^2}{2}\int_p\,\frac{n(\varepsilon_p)}{\varepsilon_p}\,,
\eeq
where $n(\varepsilon_p)=1/({\rm e}^{\beta \varepsilon_p}-1)$ and
$\varepsilon_p=\sqrt{p^2+m^2+\Pi}$.  Eq.~(\ref{tildePi1}) involves a
temperature dependent counterterm.
However the same manipulation as above,
with
$\tilde
\Pi_0$ replaced by $\tilde \Pi_0+\tilde \Pi_1$ in
Eq.~(\ref{tildePi0tad}), and the use of  the $T=0$ counterterms
which are calculated entirely from $D_0$,  eliminate it, leaving a
finite gap equation.   In
the mass-shell subtraction scheme  where
$m$ is the physical mass and the vacuum sector is trivial ($\Pi=0$),
this equation is simply:
\beq\label{PifiniteT}
\Pi\,=\,
\frac{\Gamma}{2}\int_P
D_r(P)+\frac{\Gamma}{2}\int_p\,\frac{n(\varepsilon_p)}{\varepsilon_p}.
\eeq

Consider now the 2-loop skeleton (the ``sunset''), and the
corresponding gap equation:
\beq\label{eq:gapA}
   &&\!\!\Pi(K)\!=\!-\frac{g^4}{6}\int_P\int_Q
D(P)D(Q)D(K+P+Q)\qquad\nonumber\\ &\!+\!&
\frac{\Delta g^2\!+\!\delta g^2}{2}\int_P D(P)+\Delta m^2+\delta
m^2+K^2\delta Z.
\eeq
That this expression can be made finite with the indicated
counterterms follows from a standard analysis: the
counterterm $\Delta g^2+\delta g^2$ cancels the subdivergences, while
$\Delta m^2+\delta m^2$ and $\delta Z$ cancel
the remaining global divergences (the reason behind the special
writing of the counterterms will become clear shortly).
The argument assumes, in agreement with Weinberg's theorem,  that the
repeated insertions of the self-energy in the
propagators, as generated by iterating the gap equation, do not
change in an essential way the  asymptotic form of
these propagators, expected to be typically of the form:
$\Pi(K)\simeq  K^2 F(\ln K)$ for  $K\gg m$. Note that the coupling
constant counterterm enters only a one-loop
diagram: at this order of the skeleton expansion, there is no
renormalization of the vertices of
the sunset diagram. Such renormalizations would involve skeletons
whose lowest perturbative order is $g^6$.

\begin{figure}
\includegraphics[width=5cm]{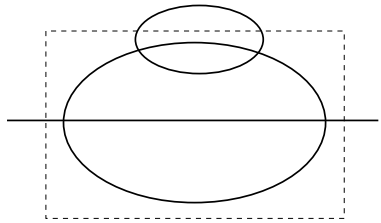}
\caption{The sunset diagram with one sunset inserted on one of the
propagator. The subdivergence contained in the
dashed line box is removed by a counterterm determined by the
Bethe-Salpeter equation.  }
\end{figure}

To  proceed to the renormalization of the BS equation we need to
take into account the fact that the  self-energy $\Pi(K)$ modifies
the asymptotic behavior
of the propagator, as indicated above.  We  then write
$\Pi=\Pi_2+\Pi_0$, where $\Pi_2(K)$ is finite and contains the
exact asymptotic  behavior of $\Pi(K)$, and $\Pi_0(K)$ grows at most
logarithmically at large $K$ ($\Pi_2(K)$ and
$\delta Z$ can be obtained by solving the gap equation with  $m=0$).
We set:
\beq
D_{-2}(K)=(K^2+m^2+\Pi_2)^{-1}.
\eeq
$D_{-2}$ will play here the  role of $D_{0}$ in the one-loop example
(note that $D_{-2}$ takes care of field
renormalization). Thus, we define an auxiliary
   4-point function $\Gamma_0$ as the solution to the BS equation
(\ref{BS1}) with $D_{-2}$ as propagators. The equation
for
$\Gamma_0(K,P)$ contains all the divergences of that for the full
4-point function
$\Gamma(K,P)$, and the renormalizations of $\Gamma_0$ and $\Gamma$
involve therefore the same counterterms. To
determine these, we first write the kernel of the BS equation as
$\Lambda_0(K,P)+\delta g^2$, with:
\beq\label{lambdasunset}
\Lambda_0(K,P)\!=\!\Delta g^2\!-\!g^4\int_Q D_{-2}(Q)D_{-2}(K+P+Q),
\eeq
and $\Delta g^2$ is chosen  so as to make $\Lambda_0(K,P)$ finite.
The counterterm  $\delta g^2$ is then adjusted, as
in Eq.~(\ref{BStad}), so as to eliminate the divergence of the equation:
\beq\label{eq:Gamma_ren_condition}
\Gamma_0(0,0)&\!=\!& \delta g^2+\Lambda_0(0,0)\nonumber\\
&\!-\!&\frac{1}{2}\int_P \Gamma_0(0,P)D_{-2}^2(P)[\delta
g^2+\Lambda_0(P,0)] ,\,\,\,\,\,\,\,\,\,
\eeq
where $\Gamma_0(0,0)$ is fixed by a renormalization condition, and
$\Gamma_0(0,P)$ can be obtained from
the following finite equation :
\begin{eqnarray}\label{eqG}
\!\!\!\!\Gamma_0(0,P)&\!\!-\!\!&\Gamma_0(0,0) \!=\!
\Lambda_0(0,P)\!-\!\Lambda_0(0,0)\nonumber\\
&\!\!-\!\!&\!\!\int_Q
\Gamma_0(0,Q)D_{-2}^2(Q)\left\{\Lambda_0(Q,P)\!-\!\Lambda_0(Q,0)\right\}.\,\,\,\,\,
\end{eqnarray}
($\Lambda_0(Q,P)-\Lambda_0(Q,0)\sim 1/Q^2$
for $Q^2\gg P^2$, so that the   integral over $Q$ is indeed finite.)

By combining
Eqs.~(\ref{eqG}) and (\ref{eq:Gamma_ren_condition}) one gets
$\Gamma_0(0,P)$ in terms of $\Lambda_0(Q,P)$ and $\delta g^2$. The
result is in fact nothing but Eq.~(\ref{BS1}) with
$K=0$, $D$ replaced by $D_{-2}$ and $\Lambda(0,P)$ replaced by
$\Lambda_0(0,P)+\delta g^2$.  We shall now use this
expression of $\Gamma_0(0,P)$ to eliminate  the vertex subdivergences from the
gap equation. To proceed, we write again
$D=D_{-2}+\delta D$, where
$\delta D=D_{-2}[-\Pi_0]D_{-2}
+D_{r}$ contains all the dependence on $\Pi_0$, and call $\tilde
\Pi_2$  the r.h.s. of Eq.~(\ref{eq:gapA})
evaluated  with
$D$ replaced by
$D_{-2}$.    Furthermore, we set
   $\tilde{\Pi}_0(K)=(1/2)\int_P [\Lambda_0(K,P)+\delta g^2]\delta
D(P)$. The gap equation is
$\Pi=\tilde
\Pi_2+\tilde\Pi_0+\tilde\Pi_r$ where  $\tilde\Pi_r(K)$ is finite and
goes as $1/K^2$ at large $K$. We then write
$\tilde{\Pi}_0(K)=\tilde{\Pi}_0(K)-\tilde{\Pi}_0(0)+\tilde{\Pi}_0(K)$, where
$\tilde{\Pi}_0(K)-\tilde{\Pi}_0(0)$ is finite, and we  express
$\Lambda_0(0,P)+\delta g^2$ in the defining equation for $\tilde
\Pi_0(0)$ in terms of $\Gamma_0(0,P)$. We get:
\beq
\tilde{\Pi}_0(0)&\!\!=\!\!&\frac{1}{2}\int_P\Gamma_0(0,P)\left[\delta D(P)+
D_{-2}^2(P)\tilde\Pi_0(P)\right].\,\,\,\,\,\,\,\,
\eeq
For $\Pi$ solution of the gap equation, we can set
$\tilde{\Pi}_0=\Pi-\tilde\Pi_2-\tilde\Pi_r$ in the r.h.s., and verify
that the divergent terms linear in $\Pi_0$
($=\Pi-\Pi_2$)  cancel, as anticipated.   Using the resulting
expression of $\tilde\Pi_0$ we obtain the solution of the
gap equation for $K=0$ in the form:
\beq\label{eq:gap_ren}
\Pi(0) &\!\! =\!\! &\frac{1}{2}\int_P
\Gamma(0,P)\left\{\Pi_2(P)\!-\!\tilde{\Pi}_2(P)\!-\!\Pi_{r}(P)\right\}D_{-2}^2(P)\nonumber\\
   & \!\!+\!\!&\frac{1}{2}
    \int_P \Gamma(0,P) D_{r}(P)+\Pi_{r}(0)+\tilde{\Pi}_{2}(0).
\eeq
To isolate the remaining divergences, we write
\beq\label{Pi2p}
\tilde\Pi_2(K)=\tilde\Pi_2'(K) + \delta m^2 +\frac{\delta
g^2}{2}\int_P D_{-2}^2(P)
\eeq
where $\tilde\Pi_2'(K)$ is finite (owing to the counterterms $\Delta
m^2$ and $\Delta g^2$). The difference
$\tilde\Pi_2'(K)-\Pi_2(K)$ is logarithmic at large $K$ and
contributes to a divergence of the first integral of
Eq.~(\ref{eq:gap_ren}).  But neither this divergence, nor those
coming from the counterterms displayed in
Eq.~(\ref{Pi2p}), depend on the solution $\Pi_0$ of the gap equation,
and they  can be absorbed in
the mas counterterm
$\delta m^2$. This completes the determination of the counterterms
which, as we have seen, can all be calculated from
$D_{-2}$.

At this point we emphasize a special feature of $\Phi$-derivable
approximations: As we have indicated earlier, the
renormalization of the two-loop skeleton generates a coupling
constant counterterm for  the one-loop skeleton, but not
for its own vertices. This is a general feature, which persists in
higher orders in the loop-expansion of $\Phi$.
Correspondingly, the
$\beta$-function deviates from
that given by perturbation theory beyond the perturbative orders
explicitly included in the skeletons
\cite{BIU03}. For instance, in the present example, the perturbative
$\beta$-function is correctly reproduced to order
$g^4$ (when one adds the two contributions of the tadpole and sunset
diagrams), but deviates at order $g^6$.

The extension of the previous analysis to finite temperature brings
no new ultraviolet difficulty.
Again, we can  separate each loop integral in the sunset into a ``vacuum''
contribution, and a contribution containing a statistical
factor.  The final expression for the self-energy takes then a form
similar to that at zero
temperature, and may be written as
$\Pi=\tilde\Pi_2+\tilde\Pi_r+\tilde\Pi_0+\tilde\Pi_1+\tilde\Pi_3$.
The first
contribution, $\tilde\Pi_2$, is the same as before and does not
depend on the temperature. The last contribution,
$\tilde\Pi_3(K)$, is one in which each of the loop integrals contains
a statistical factor. It  is finite and decreases
as
$1/K^2$ at large $K$; thus it is not involved in any divergent term,
and  it can be regarded as a simple correction to
$\tilde\Pi_r$. Finally,  $\tilde \Pi_0$ is defined as at zero
temperature, and \cite{BIU03}:
\beq
\tilde\Pi_1(K)=\frac{1}{2}\int_{p_0,p}\left[\Lambda_0(K,P)+\delta
g^2\right]\rho(p_0,p)n_{|p_0|}\sigma_{p_0}
\eeq
where $\rho(p_0,p)$ is the spectral function of the propagator $D$,
$\sigma_{p_0}$
denotes the sign of
$p_0$, and the integral runs over the real $p_0$ axis. As in the
one-loop example, we can combine
$\tilde
\Pi_1$ with $\tilde \Pi_0$  and show that the zero temperature
coupling constant counterterms eliminate the
apparent divergence depending on the temperature.

We are grateful to H. van Hees and J. Knoll for stimulating
discussions concerning their work, and to A. Rebhan for useful comments
on this manuscript.

\end{document}